\documentstyle[manuscript,aps]{revtex}
\begin{document}
\draft
\title{Do gravitational waves carry energy-momentum and angular momentum?}
\author{Janusz Garecki}
\address{Institute of Physics, University of Szczecin, Wielkopolska 15, 70-451
Szczecin, POLAND}
\date{\today}
\maketitle
\begin{abstract}
In the paper we show that the real gravitational waves which have
$R_{iklm}\not= 0$ always carry energy-momentum and angular momentum. Our proof
uses canonical superenergy and supermomentum tensors for gravitational field.
\end{abstract}
KEY WORDS: gravitational waves, gravitational energy, gravitational
superenergy 
\pacs{04.20.Me.04.30.+x}
\newpage
\section{Introduction}
In General Relativity ({\bf GR}) the gravitational field $\Gamma^i_{kl}$ does
not possess any energy-momentum tensor but it only possesses the so-called
``energy-momentum pseudotensors''. It is a consequence of the Einstein
Equivalence Principle ({\bf EEP}). So, many authors [1,4,8,18] doubt in reality
of the energy-momentum and angular momentum transfer by gravitational waves. As
an argumentation some of these authors use the fact that for the majority exact
solutions of the vacuum Einstein equations which, as we think, can represent
gravitational waves, pseudotensors {\it globally vanish} in some coordinates. In consequence, they give in these coordinates
``no gravitational energy and no gravitational energy flux''. Other of these authors [1,8]
try to use vanishing of the components $_g t^{ok}$ (or $_g t_0^{~ k}$)
of a gravitational energy-momentum pseudotensor $_g t^{ik}$ (or $_g t_i^{~ k}$) as
a coordinate condition coupled to the Einstein equations and then also get
(in special coordinates), as they assert, ``global vanishing of the pure
gravitational energy and gravitational energy flux''. 

However, {\it such conclusions are physically uncorrect}. Firstly, these
authors neglect an important role  of the four-velocity $\vec v,~ ~ \vec
v\cdot\vec v = 1$ of an observer {\bf O}, which is measuring gravitational (or
other) field, in definition of the energy density $\epsilon$ and energy flux
$P^i$ of the field. Namely, the correct definitions of the energy density and
its flux for such observer are the following
\begin{equation}
\epsilon = T^{ik}v_iv_k = T_{ik}v^iv^k,
\end{equation}
\begin{equation}
P^i = \bigl(\delta^i_k - v^i v_k\bigr)T^{kl}v_l;
\end{equation}
not
\begin{equation}
\epsilon = T^{00}, ~ ~ P^{\alpha} = T^{0\alpha}.
\end{equation}
$i,k,l,..., = 0,1,2,3; ~ ~ \alpha,\beta,\gamma,\delta,..., = 1,2,3$. $T^{ik}$
mean here the components of an energy-momentum tensor.

$T^{00}$ and $T^{0\alpha}$ give energy density $\epsilon$ and energy flux $P^i$
for the observer {\bf O} iff in the used global coordinates we have for the
observer $v^i = \delta^i_0,~ ~ v_k = \delta^0_k$.\footnote{This means
physically that the observer {\bf O} is at rest with respect to the used
coordinates.} 

The above facts are also true for a gravitational energy-momentum pseudotensor
$_g t^{ik}$ (or $_g t_i^{~ k}$). 

In consequence, even if $_g t^{0k} = 0$ (or $_g t_0^{~k} = 0$) globally in the
used coordinates then, as one can easily see, {\it not for all observers}
$\epsilon = P^i = 0$. All depends on the four-velocity $\vec v$ of the observer
{\bf O}.  

Moreover, one coordinate condition of the kind $_g t^{0k} = 0$ (or $_g
t_0^{~k} = 0$) {\it does not suffice for the all variety of the gravitational
energy-momentum pseudotensors}. In fact, we need here an own coordinate
condition for every gravitational energy-momentum pseudotensor.

Secondly, and it is the most important fault, these authors forget that
gravitational energy-momentum (and gravitationam angular momentum)
pseudotensors, as being functions of th Levi-Civita connection coefficients
\footnote{Physically, this connection plays the role of the total gravitational
strenghts.} describe energy-momentum of the {\it total gravitational field}
which is a combination of the real gravitational field (for which
$R_{iklm}\not= 0$) and inertial forces field (for which $R_{iklm} = 0$). The
inertial forces field is generated by the used coordinates. This is also a
consequence of the {\bf EEP}. 

So, if in some coordinates $_g t^{ik}$ or ($_g t_i^{~ k}$) (or only $_g
t^{ok}$ or $_g t_0^{~ k}$) globally vanish, this does not mean that in
these coordinates we have no pure gravitational energy and no pure
gravitational energy flux.\footnote{Even if we confine to the observers which
are at rest in the used global coordinates and for which $_g t^{00}$ (or$_g
t_0^{~ 0}$) is ``energy density'' and $_g t^{0\alpha}$ (or $_g t_0^{~ \alpha}$)
give the ``energy flux''.} Simply, it means only that in such coordinates {\it
energy and energy flux of the real gravitational field cancel} with the energy
and energy flux of the  inertial forces field.\footnote{Energy-momentum of the
inertial forces field gives contribution to energy-momentum pseudotensors.} Energy and energy-momentum flux of the
real gravitational field which has $R_{iklm} \not= 0$ {\it always exist and
are non-null}. In order to show this very important fact one can use {\it
the canonical superenergy and canonical angular supermomentum
tensors}\footnote{Or other superenergy and angular supermomentum tensors.} for gravitational
field [10-14]. 

The canonical superenergy tensor $_g S_i^{~k}$ and canonical angular
supermomentum tensor $_g S^{ikl} = (-) _g S^{kil}$ are constructed in
such a way that they {\it extract covariant information} about real
gravitational field which is hidden in canonical energy-momentum and
canonical angular momentum pseudotensors. Namely, these superenergy and
angular supermomentum tensors are obtained from suitable gravitational
pseudotensors {\it by some kind of averaging} and they are
functions of the curvature tensor and its covariant derivatives only.
So, they {\it really describe only real gravitational field}, which has $R^i_{~klm}\not=
0$.

The paper is organized as follows. In Section II we remember canonical
superenergy and supermomentum tensors for gravitational field in {\bf
GR}. In Section III we give a summary about applications of the
canonical superenergy and supermomentum tensors for gravitational field
to analysis of the gravitational waves and in Section IV we give some
concluding remarks.
\section{Constructive definition of the canonical superenergy tensor and
canonical angular supermomentum tensor}
We remember here the constructive, general definition of the superenergy
tensor $S_a^{~b}(P)$ applicable to gravitational field and to matter
fields. 

In the normal coordinates {\bf NC(P)} [see, e.g., 20,21,22], we define
\begin{equation}
S_{(a)}^{~~~(b)}(P) := (-)\displaystyle\lim_{\Omega\to
P}{\int\limits_{\Omega}\biggl[T_{(a)}^{~~~(b)}(y) -
T_{(a)}^{~~~(b)}(P)\biggr]d\Omega\over
1/2\int\limits_{\Omega}\sigma(P;y)d\Omega},
\end{equation}
where
\begin{equation}
T_{(a)}^{~~~(b)}(y) := T_i^{~k}(y)e^i_{(a)}(y)e^{(b)}_k(y),
\end{equation}
\begin{equation}
T_{(a)}^{~~~(b)}(P):= T_i^{~k}(P)e^i_{(a)}e^{(b)}_k(P) = T_a^{~b}(P)
\end{equation}
are the so-called {\it physical or tetrad components} of the
pseudotensor (or tensor) field $T_i^{~k}(y)$ which describes
energy-momentum. $\{y^i\}$ are the normal coordinates. $e^i_{(a)}(y),
~~e^{(b)}_k(y)$ mean here an orthonormal tetrad $e^i_{~(a)}(P) =
\delta^i_a$ and its dual $e^{(a)}_k(P) = \delta^a_k$ parallelly
propagated along geodesics throught $P$ ($P$ = origin of the {\bf
NC(P)}). 

We have 
\begin{equation}
e^i_{(a)}(y) e_i^{(b)}(y) = \delta_a^b.
\end{equation}
$\Omega$ is a sufficiently small ball 
\begin{equation}
y^{0^2} + y^{1^2} + y^{2^2} + y^{3^2} \leq R^2
\end{equation}
for an auxilliary positive-definite metric $h^{ik} := 2 v^iv^k -
g^{ik}$, i.e., $\Omega$ can be given as 
\begin{equation}
h_{ik}y^iy^k\leq R^2.
\end{equation}
$v^i$ denotes here the components of the four-velocity $\vec v : v^lv_l
= 1$ of an observer {\bf O} which is at rest at the beginning {\bf P} of
the used normal coordinates {\bf NC(P)}. 

The ball $\Omega$ surrounds {\bf P}: {\bf P} is centre of the ball.
\begin{equation}
\sigma(P;y) {\dot =} {1\over 2}\bigl(y^{0^2} - y^{1^2} - y^{2^2} -
y^{3^2}\bigr)
\end{equation}
is the two-points {\it world function} which has been introduced by J.L.
Synge years ago [23]. The symbol ${\dot =}$ means that an equation is
only valid for special coordinates. The latter can be covariantly
defined by the eikonal-like equation
\begin {equation}
g^{ik}\partial_i\sigma\partial_k\sigma = 2\sigma
\end{equation}
together with $\sigma(P,P) = 0, ~~\partial_i\sigma(P,P) = 0$. 

The $\Omega$ can be also given by the inequality
\begin{equation}
h^{ik}\partial_i\sigma\partial_k\sigma\leq R^2.
\end{equation} 

Since the tetrad components and normal components are equal at the point
{\bf P}, we will write the components of any quantity attached to the
point {\bf P} without tetrad brackets , e.g., we will write
$S_a^{~b}(P)$ instead of $S_{(a)}^{~~~(b)}(P)$ and so on.

If $T_i^{~k}$ are the components of a symmetric energy-momentum tensor
of matter then we will get from (4) 
\begin{equation}
_m S_a^{~b}(P) = \delta^{mn}\nabla_{(m}\nabla{_n)} {\hat T}_a^{~b}.
\end{equation}
Hat over a quantity denotes its value at {\bf P}.

By using of the four-velocity $\vec v$ of a fictitious observer {\bf O}
being at rest at the beginning {\bf P} of the used {\bf NC(P)} and local
metric ${\hat g}^{ab} {\hat =}  \eta^{ab}$ one can write this covariantly as 
\begin{equation}
_m S_a^{~b} (P;v^l) = \bigl(2{\hat v}^l{\hat v}^m - {\hat
g}^{lm}\bigr)\nabla_{(l}\nabla_{m)}{\hat T}_a^{~b}.
\end{equation}
The last formula gives us {\it the canonical superenergy tensor for
matter}. 

For the gravitational field $\Gamma^i_{kl} = \Bigl\{^i_{kl}\Bigr\}$, after
substituting into (4)  $T_i^{~k} = _E t_i^{~k}$, where 
\begin{eqnarray}
_E t_i^{~k} & = & {c^4\over 16\pi G}\Bigl\{\delta^k_i
g^{ms}\bigl(\Gamma^l_{mr}\Gamma^r_{sl} -
\Gamma^r_{ms}\Gamma^l_{rl}\bigr)\nonumber \\
& + &  g^{ms}_{~~,i}\bigl[\Gamma^k_{ms} -{1\over
2}\bigl(\Gamma^k_{tp}g^{tp} - \Gamma^l_{tl}g^{kt}\bigr)g_{ms}\nonumber
\\
& - & {1\over 2}\bigl(\delta^k_s \Gamma^l_{ml} + \delta^k_m
\Gamma^l_{sl}\bigr)\bigr]\Bigr\}
\end{eqnarray}
is the {\it canonical Einstein energy-momentum pseudotensor} for the
gravitational field, we will obtain
\begin{equation}
_g S_a^{~b}(P;v^l) = \bigl(2{\hat v}^l{\hat v}^m - {\hat g}^{lm}\bigr)
{_E{\hat T}_a^{~b}}{}_{lm}.
\end{equation}
\begin{eqnarray}
{_E T_a^{~b}}{}_{lm} & = & {2\alpha\over 9}\biggl[B^b_{~alm} +
P^b_{~alm}\nonumber \\
& - & {1\over 2}\delta^b_a R^{ijk}_{~~~m}\bigl(R_{ijkl} + R_{ikjl}\bigr)
+ 2\delta^b_a\beta^2 E_{(l\vert g}{}E^g_{~\vert m)}\nonumber \\
& - & 3\beta^2 E_{a(l\vert}{}E^b_{~\vert m)} + 2\beta
R^b_{~(ag)(l\vert}{}E^g_{~\vert m)}\biggr].
\end{eqnarray}

In the last formula 
\begin{equation}
\alpha = {c^4\over 16\pi G} = {1\over 2\beta}
\end{equation}
and 
\begin{equation}
E^k_i := T^k_i - {1\over 2}\delta^k_i T
\end{equation}
is the modified energy-momentum of matter. On the other hand 
\begin{equation}
B^b_{~alm} := 2R^{bik}_{~~~(l\vert}{}R_{aik\vert m)} -{1\over
2}\delta^b_a R^{ijk}_{~~~l}{} R_{ijkm}
\end{equation}
are the components of the {\it Bel-Robinson} tensor, while the tensor 
\begin{equation}
P^b_{~alm} := 2R^{bik}_{~~~(l\vert}{}R_{aki\vert m)} - {1\over
2}\delta^b_a R^{ijk}_{~~~l}{}R_{ikjm}
\end{equation}
is closely related to the latter. 

The tensor $_g S_a^{~b}(P;v^l)$ is the {\it canonical superenergy
tensor} for gravitational field.\footnote{We must emphasize that the
canonical superenergy tensor $_g S_a^{~b}(P;v^L)$ is originally a
function of the curvature components $R_{iklm}$ and tensor Ricci components
$R_{ik}$. We have eliminated the all Ricci components by use of
the Einstein equations $R_{ik} = \beta E_{ik}$.}

In vacuum the tensor $_g S_a^{~b}(P;v^l)$ takes the simpler form 
\begin{eqnarray}
_g S_a^{~b}(P;v^l) & = & {8\alpha\over 9}\bigl(2{\hat v}^l{\hat v}^m -
{\hat g}^{lm}\bigr)\bigl[{\hat R}^{b(ik)}_{~~~~~(l\vert}{}{\hat
R}_{aik\vert m)}\nonumber \\
& - & {1\over 2}\delta^b_a{\hat R}^{i(kp)}_{~~~~~(l\vert}{}{\hat
R}_{ikp\vert m)}\bigr]
\end{eqnarray}
and the quadratic form $_g S_{ab}(P;v^l)v^av^b$, where $v^av_a = 1$,
{\it is positive-definite}.

The canonical angular supermomentum tensor we define in analogy to the
definition of the canonical superenergy tensor. Namely, we define in the
normal coordinates {\bf NC(P)} 
\begin{equation}
S^{(a)(b)(c)}(P) = S^{abc}(P) := (-)\displaystyle\lim_{\Omega\to P}{\int\limits_{\Omega}
\biggl[M^{(a)(b)(c)}(y) - M^{(a)(b)(c)}(P)\biggr]d\Omega\over
1/2\int\limits_{\Omega}\sigma(P;y)d\Omega},
\end{equation}
where, as formerly,
\begin{equation}
M^{(a)(b)(c)}(y) := M^{ikl}(y)e^{(a)}_i(y) e^{(b)}_k(y) e^{(c)}_l(y),
\end{equation}
\begin{eqnarray}
M^{(a)(b)(c)}(P) & :=&  M^{ikl}(P)e^{(a)}_i(P) e^{(b)}_k(P)
e^{(c)}_l(P)\nonumber \\
& = & M^{ikl}(P)\delta^a_i\delta^b_k\delta^c_l = M^{abc}(P) 
\end{eqnarray}
are {\it physical} (or tetrad) components of the field $M^{ikl}(y) = (-)
M^{kil}(y)$ describing angular momentum. 

For matter, as $M^{ikl}(y) = (-) M^{kil}(y)$, we take 
\begin{equation}
M^{ikl}(y) = \sqrt{\vert g\vert}\bigl(y^i T^{kl} - y^k T^{il}\bigr),
\end{equation}
where $T^{ik} = T^{ki}$ are the components of the  symmetric
energy-momentum tensor of matter\footnote{This tensor is the source in
the Einstein equations.} and $\bigl\{y^i\bigr\}$ denote normal
coordinates.

The expression (26) gives us the {\it total angular momentum densities},
orbital and spin, because the dynamical tensor $T^{ik} = T^{ki}$ is
obtained from canonical one by means of the Belinfante symmetrization
procedure (and, therefore, includes material spin). 

For gravitational field, as $_g M^{ikl}(y)$,  we favorize and take the expression
most closely related to the Einstein canonical energy-momentum complex (See Appendix). We
will call this expression {\it canonical} also. Namely, as $_g M^{ikl}(y)$, we
take the expression given by Bergmann and Thomson [24]
\begin{equation}
_g M^{ikl}(y) = _F U^{i[kl]}(y) - _F U^{k[il]}(y) + \sqrt{\vert
g\vert}\bigl(y^i _{BT} t^{kl} - y^k _{BT} t^{il}\bigr).
\end{equation}
In the last formula
\begin{equation}
_F U^{i[kl]} := g^{im} {}_F U_m^{~[kl]},
\end{equation}
where $_F U_m^{~[kl]}$ mean {\it von Freud superpotentials} and
\begin{equation}
_{BT} t^{kl} := g^{ki} _E t_i^{~l} + {g^{mk}_{~~,p}\over\sqrt{\vert
g\vert}}{} _F U_m^{~~[lp]}\end{equation}
is the {\it Bergmann-Thomson gravitational energy-momentum
pseudotensor} [24].

One can interpret the Bergmann-Thomson expression (27) as a sum of the
{\it spinorial part} 
\begin{equation}
S^{ikl} := _F U^{i[kl]} - _F U^{k[il]}
\end{equation}
and {\it orbital part}
\begin{equation}
O^{ikl} := \sqrt{\vert g\vert}\bigl(y^i _{BT} t^{kl} - y^k _{BT}
t^{il}\bigr)
\end{equation}
of the gravitational angular momentum densities.

Substituting (26) into (23) we will get the {\it canonical angular
supermomentum tensor for matter}
\begin{eqnarray}
_m S^{abc}(P;v^l) & = & 2\bigl[\bigl(2{\hat v}^a{\hat v}^p - {\hat
g}^{ap}\bigr) \nabla_p{\hat T}^{bc}\nonumber \\ & - & \bigl(2{\hat
v}^b{\hat v}^p - {\hat g}^{bp}\bigr)\nabla_p{\hat T}^{ac}\bigr].
\end{eqnarray}

If we substitute (27) into (23), then we will obtain the {\it
gravitational canonical angular supermomentum tensor}\footnote{We have
also eliminated in (33) the components of Ricci tensor with the help of
the Einstein equations.}
\begin{eqnarray}
_g S^{abc}(P;v^l) & = & \alpha\bigl(2{\hat v}^p{\hat v}^t - {\hat
g}^{pt}\bigr) \bigl[\beta\bigl({\hat g}^{ac}{\hat g}^{br} - {\hat
g}^{bc}{\hat g}^{ar}\bigr)\nabla_{(t}{}{\hat E}_{pr)}\nonumber \\ & + &
2{\hat g}^{ar}\nabla_{(t} {\hat R}^{(b}_{~~p}{}^{c)}_{~~r)} - 2{\hat
g}^{br}\nabla_{(t} {\hat R}^{(a}_{~~p}{}^{c)}_{~~r)}\nonumber \\ & + &
2/3{\hat g}^{bc}\bigl(\nabla_r {\hat R}^r_{~(t}{}^a_{~p)} -
\beta\nabla_{(p} {\hat E}^a_{t)}\bigr) \nonumber \\ & - & 2/3 {\hat
g}^{ac}\bigl(\nabla_r {\hat R}^r_{~(t}{}^b_{~p)} - \beta\nabla_{(p}
{\hat E}^b_{t)}\bigr)\bigr].
\end{eqnarray}

In vacuum, i.e., when $T_{ik} = 0 \equiv E_{ik} := T_{ik} - 1/2 g_{ik}
T = 0$, the gravitational canonical angular supermomentum tensor $_g
S^{abc}(P;v^l) = (-) _g S^{bac}(P;v^l)$ simplifies to the form
\begin{eqnarray}
_g S^{abc}(P;v^l) & = & 2\alpha\bigl(2{\hat v}^p{\hat v}^t - {\hat
g}^{pt}\bigr)\bigl[{\hat g}^{ar} \nabla_{(p} {\hat
R}^{(b}_{~~t}{}^{c)}_{~~r)}\nonumber \\ & - & {\hat g}^{br} \nabla_{(p}
{\hat R}^{(a}_{~~t}{}^{c)}_{~~r)}\bigr].
\end{eqnarray}

It is interesting that the orbital part $0^{ikl} = \sqrt{\vert
g\vert}\bigl(y^i _{BT} t^{kl} - y^k _{BT} t^{il}\bigr)$ of the $_g
M^{ikl}$ gives {\it no contribution} to the tensor $_g S^{abc}(P;v^l)$.
Only spin part $S^{ikl} = _F U^{i[kl]} - _F U^{k[il]}$ {\it gives
non-zero contribution} to this tensor. Notice also that the canonical
angular supermomentum tensors $_g S^{abc}(P;v^l)$ and $_m
S^{abc}(P;v^l)$, gravitation and matter, {\it needn't any radius vector}
for their own existing. 
\section{Summary about application of the superenergy and angular supermomentum
tensors to analysis of the gravitational waves}
One can easily check by a direct calculation that the canonical
superenergy tensor $_g S_i^{~k}(P;v^l)$ and the canonical angular
supermomentum tensor $_g S^{ikl}(P;v^m)$ give {\it positive-definite}
superenergy density, {\it non-vanishing} superenergy  flux and {\it
non-vanishing} angular supermomentum flux for every known solution to
the vacuum Einstein equations which represents a gravitational wave with
$R_{iklm}\not= 0$. This is true in any admissible coordinates and it all
was showed in our papers [10,12--14]. It results from the above fact
that every real gravitational wave, which has $R_{iklm}\not= 0$, {\it
must carry} gravitational energy-momentum and gravitational angular
momentum. If not, then there would be a {\it contradiction} between
``energy level'' and ``superenergy level'' because the canonical superenergy
and angular supermomentum tensors  originated by some kind of averaging
of the suitable pseudotensors. Notice also in this context that, as it
follows from the definition (4), the $\epsilon(\Omega) := (-)
_g S_{ik}(P;v^l)v^iv^k\int\limits_{\Omega}\sigma(P;y)d\Omega$ gives the
approximate (relative with respect {\bf P}) energy contained in the
sufficiently small domain $\Omega$ and $P^i(\Omega) := (-)1/2\bigl(\delta^i_k
- u^iu_k\bigr)_g S_l^{~k}(P;v^m)v^l\int\limits_{\Omega}\sigma(P;y)d\Omega$ 
gives Poynting's vector of the (relative with respect {\bf P}) energy
flux contained in $\Omega$. $\epsilon(\Omega)$ and $P^i(\Omega)$
{\it do not vanish} in any admissible coordinates for a real
gravitational wave which has $R_{iklm}\not= 0$. 

So, the non-zero superenergy density and its non-zero flux really {\it
demand non-zero gravitational energy  and its non-zero flux}. 

On the ``energy-momentum level'', where we use pseudotensors, the very
important fact that any gravitational wave which has $R_{iklm}\not= 0$
{\it always transfer energy-momentum} is
camouflaged in some coordinates by energy-momentum of the inertial forces
field. 
 
The analogical considerations and conclusion  which are based on the definition
(23) are valid also for the angular momentum.  
\section{Concluding remarks}
If you want to get the correct information about energy-momentum and
angular momentum of the real gravitational field by use pseudotensors,
then you must use the pseudotensors in very special coordinates only
(see eg. [25-27]).
Namely, you must use pseudotensors in a coordinates in which
$\Gamma^i_{kl}$ describe only the real gravitational field. The examples
of such coordinates are given, e.g., by {\it global Bondi-Sachs 
coordinates} for a closed system [7] or, in general, by normal
coordinates {\bf NC(P)} [20-22, 28,29].

In order to get information on gravitational energy-momentum and angular
momentum in arbitrary admissible coordinates one must use covariant
expressions which depend on curvature tensor. Our canonical superenergy
and angular supermomentum tensors are exactly the quantities of such a
kind. In application to gravitational radiation these quasilocal
quantities unambiguously show that any gravitational waves which has
$R_{iklm}\not= 0 $ always transfer energy-momentum and angular momentum.
So, the conclusions given in the papers like [1,8,18], where authors use pseudotensors only
{\it are uncorrect}.\footnote{In the paper [18] the Author uses also
{\it stationary Tolman integral} to analysis of a dynamical system. Of course,
this is {\it unjustified physically}.}
\appendix
\section{The canonical energy-momentum complex in General Relativity}
One can easily transform \footnote{ It is the most easily to do by use the
formalism of the exterior differential forms.} Einstein equations to the
superpotential form 
\begin{equation}
\sqrt{\vert g\vert}\bigl(T_i^k + _E t_i^{~k}\bigr) = {_F U_i^{~[kl]}}_{,l}.
\end{equation}
In the last formula $T^{ik} = T^{ki}$ is the symmetric energy-momentum
tensor for matter, $_E t_i^{~k}$ mean the components of the {\it Einstein
canonical energy-momentum pseudotensor for gravitational field} and $_F
U_i^{~[kl]}$ denote {\it von Freud superpotentials}. The sum 
\begin{equation}
\sqrt{\vert g\vert}\bigl(T^k_i + _E t_i^{~k}\bigr) =: _E K_i^{~k}
\end{equation}
is called {\it the canonical Einstein energy-momentum complex for matter
and gravitation in General Relativity} and it is usually denoted by $_E
K_i^{~k}$.

In extended form we have
\begin{eqnarray}
_E t_i^{~k} & = & {c^4\over 16\pi G}\biggl\{\delta_i^k
g^{ms}\bigl(\Gamma^l_{mr}\Gamma^r_{sl} -
\Gamma^r_{ms}\Gamma^l_{rl}\bigr)\nonumber \\ & + &
g^{ms}_{~~,i}\bigl[\Gamma^k_{ms} - 1/2\bigl(\Gamma^k_{tp} g^{tp} -
\Gamma^l_{tl} g^{kt}\bigr)g_{ms}\nonumber \\ & - & 1/2\bigl(\delta^k_s
\Gamma^l_{ml} + \delta^k_m \Gamma^l_{sl}\bigr)\bigr]\biggr\},
\end{eqnarray}
\begin{equation}
_F U_i^{~[kl]} = {c^4\over 16\pi G}{1\over\sqrt{\vert
g\vert}}g_{ia}\bigl[(-g)\bigl(g^{ka} g^{lb} - g^{la}
g^{kb}\bigr)\bigr]_{,b}.
\end{equation}

From (A.1) there follow  the local, differential energy-momentum conservation
laws 
\begin{equation}
\bigl[\sqrt{\vert g\vert}\bigl(T_i^{~k} + _E t_i^{~k}\bigr)\bigr]_{,k} =
0,
\end{equation}
for matter and gravitation.

\end{document}